\documentclass[12pt,letter]{article}

\usepackage{graphicx, epsfig, color}
\def\to{\rightarrow}

\def\bi{\begin{itemize}}
\def\ei{\end{itemize}}

\def\ta{\tilde a}
\def\tG{\tilde G}

\def\tst{\tilde t}

\def\tg{\tilde g}

\def\tw{\widetilde W}
\def\tz{\widetilde Z}
\def\alt{\stackrel{<}{\sim}}
\def\agt{\stackrel{>}{\sim}}
\def\be{\begin{equation}}  
\def\ee{\end{equation}}  
\newcommand\annp[3]{{\it Annals\ Phys.\ }{\bf #1} (#2) #3}
\newcommand\prd[3]{{\it Phys.\ Rev.\ }{\bf D #1} (#2) #3}

\newcommand\prl[3]{{\it Phys.\ Rev.\ Lett.\ }{\bf #1} (#2) #3}
\newcommand\plb[3]{{\it Phys.\ Lett.\ }{\bf B #1} (#2) #3}
\newcommand\njp[3]{{\it New\ J.\ Phys.\ }{\bf #1} (#2) #3}
\newcommand\jhep[3]{{\it J. High Energy Phys.\ }{\bf #1} (#2) #3}

\newcommand\ijmpa[3]{{\it Int.\ J.\ Mod.\ Phys.\ }{\bf A #1} (#2) #3}
\newcommand\npb[3]{{\it Nucl.\ Phys.\ }{\bf B #1} (#2) #3}

\newcommand\ptp[3]{{\it Prog.\ Theor.\ Phys.\ }{\bf #1} (#2) #3}

\newcommand{\hepph}[1]{hep-ph/#1}

\newcommand{\astroph}[1]{astro-ph/#1}
\newcommand\arnps[3]{{\it Ann.\ Rev.\ Nucl.\ Part.\ Sci.}{\bf  #1} (#2) #3}

\begin{document}
\begin{titlepage}
\begin{flushright}
NSF-KITP-08-15\\ 
FSU-HEP/080315
\end{flushright}

\vspace{0.5cm}
\begin{center}
{\Large \bf 
$SO(10)$ SUSY GUTs, the gravitino problem, \\
non-thermal leptogenesis and axino dark matter
}\\ 
\vspace{1.2cm} \renewcommand{\thefootnote}{\fnsymbol{footnote}}
{\large Howard Baer $^{1,2}$\footnote[1]{Email: baer@hep.fsu.edu }
and Heaya Summy$^2$\footnote[2]{Email: heaya@hep.fsu.edu }} \\
\vspace{1.2cm} \renewcommand{\thefootnote}{\arabic{footnote}}
{\it 
1. Kavli Institue for Theoretical Physics,
University of California, Santa Barbara, CA 93106-4030 USA \\
2. Dept. of Physics,
Florida State University, Tallahassee, FL 32306, USA \\
}

\end{center}

\vspace{0.5cm}
\begin{abstract}
\noindent 
Simple SUSY GUT models based on the gauge group $SO(10)$ 
require $t-b-\tau$ Yukawa coupling unification, in addition 
to gauge coupling and matter unification. The Yukawa coupling unification
places strong constraints on the expected superparticle mass spectrum,
with scalar masses $\sim 10$ TeV while gaugino masses are quite light.
A problem generic to all supergravity models comes from overproduction of
gravitinos in the early universe: if gravitinos are unstable, then their late decays
may destroy the predictions of Big Bang nucleosynthesis. We present a 
Yukawa-unified $SO(10)$ SUSY GUT scenario which avoids the gravitino problem, 
gives rise to the correct matter-antimatter asymmetry via non-thermal
leptogenesis, and is consistent with the WMAP-measured abundance of 
cold dark matter due to the presence of an axino LSP. To maintain a 
consistent cosmology for Yukawa-unified SUSY models, 
we require a re-heat temperature $T_R\sim 10^6-10^7$ GeV, an axino mass
around $\sim 0.1-10$ MeV, and a PQ breaking scale $f_a\sim 10^{12}$ GeV. 

\vspace{0.8cm}

\end{abstract}


\end{titlepage}

\section{$SO(10)$ SUSY GUTs and Yukawa unification}
\label{sec:so10}

Grand unified theories (GUTs) are amongst the most compelling ideas
in theoretical physics. Their beauty is only enhanced via a 
marriage to supersymmetry (SUSY). The $SU(5)$ theory\cite{su5} unifies the
Standard Model (SM) gauge symmetries into single Lie group, 
while explaining the ad-hoc hypercharge assignments of the SM fermions, and 
successfully predicting the $m_b/m_\tau$ ratio. Adding SUSY to the $SU(5)$
theory stabilizes the hierarchy of interactions, but also receives
experimental support from the celebrated unification of gauge 
couplings at scale $M_{GUT}\simeq 2\times 10^{16}$ GeV.

The $SO(10)$ SUSY GUT theory has even further successes\cite{so10}. 
For one, it explains the ad-hoc anomaly cancellation within the 
SM and $SU(5)$ theories. Further, it unifies
all matter of a single generation into the 16-dimensional spinor
representation $\hat{\psi}(16)$, 
provided one adds to the set of supermultiplets
a SM gauge singlet superfield $\hat{N}^c_i$ ($i=1-3$ is a generation index) 
containing a right-handed neutrino.\footnote{Here, we adopt the 
superfield ``hat'' notation as presented in Ref. \cite{wss}.}
Upon breaking of $SO(10)$, a superpotential term 
$\hat{f}\ni {1\over 2}M_{N_i}\hat{N}^c_i\hat{N}^c_i$ is induced which 
allows for a Majorana neutrino mass $M_{N_i}$ which is necessary for
implementing the see-saw mechanism for neutrino masses\cite{seesaw}. 
In addition, the $SO(10)$ theory allows for unification of Yukawa
couplings of each generation. This applies calculationally
especially to the third generation, where in simple $SO(10)$ SUSY GUTs,
we may expect $t-b-\tau$ Yukawa coupling unification in addition 
to gauge coupling unification at scale $Q=M_{GUT}$\cite{early,also}.

In spite of these impressive successes, GUTs and also SUSY GUTs have been beset
with a variety of problems, most of them arising from implementing GUT 
gauge symmetry breaking via large, unwieldy Higgs representations.
Happily, in recent years physicists have learned that GUT theories--
as formulated in spacetime dimensions greater than four-- can use
extra-dimension compactification to break the GUT symmetry instead\cite{xdguts}. 
This is much in the spirit of string theory, where anyway one must pass
from a 10 or 11 dimensional theory to a 4-d theory via some sort of
compactification. 

Regarding Yukawa coupling unification in $SO(10)$, 
the calculation begins with stipulating the $b$ and $\tau$
running masses at scale $Q=M_Z$ (for two-loop running, we adopt the 
$\overline{DR}$ regularization scheme) and the $t$-quark running 
mass at scale $Q=m_t$. The Yukawa couplings are evolved to scale $Q=M_{SUSY}$, 
where threshold corrections must be implemented\cite{hrs}, as one passes from the 
SM effective theory to the Minimal Supersymmetric Standard Model (MSSM) effective theory.
From $M_{SUSY}$ on to $M_{GUT}$, Yukawa coupling evolution is performed using two-loop
MSSM RGEs. 
Thus, Yukawa coupling unification ends up depending on the complete SUSY mass spectrum
via the $t$, $b$ and $\tau$ self-energy corrections.

In this letter, we adopt the Isajet 7.75 program for calculation of the 
SUSY mass spectrum and mixings\cite{isajet} and IsaReD\cite{isatools} for
the neutralino relic density.
Isajet uses full two-loop RG running for all gauge and
Yukawa couplings and soft SUSY breaking (SSB) terms. In running from $M_{GUT}$ 
down to $M_{weak}$, the RG-improved 1-loop effective potential is minimized at 
an optimized scale choice $Q=\sqrt{m_{\tst_L}m_{\tst_R}}$, which accounts for
leading two-loop terms. Once a tree-level SUSY/Higgs  spectrum is calculated, the
complete 1-loop corrections are calculated for all SUSY/Higgs particle masses.
Since the SUSY spectrum isn't known at the beginning of the calculation, an iterative
approach must be implemented, which stops when an appropriate convergence criterion is satisfied.

Yukawa coupling unification has been examined in a number of previous 
papers\cite{early,also, bdft,bf,bdr,bkss}.
The parameter space to be considered is given by
\be
m_{16},\ m_{10},\ M_D^2,\ m_{1/2},\ A_0,\ \tan\beta ,\ sign (\mu ) 
\label{eq:pspace}
\ee
along with the top quark mass, which we take to be $m_t=171$ GeV.
Here, $m_{16}$ is the common mass of all matter scalars at $M_{GUT}$, 
$m_{10}$ is the common Higgs soft mass at $M_{GUT}$ and $M_D^2$ 
parameterizes either $D$-term splitting (DT) or Higgs-only soft mass splitting (HS).
The latter is given by $m_{H_{u,d}}^2=m_{10}^2\mp 2M_D^2$. 
As in the minimal supergravity (mSUGRA) model, $m_{1/2}$ is a common GUT scale gaugino mass, $A_0$ is a common
GUT scale trilinear soft term, and the bilinear SSB term $B$ has been traded for 
the weak scale value of $\tan\beta$ via the EWSB minimization conditions. The latter also
determine the magnitude (but not the sign) of the superpotential Higgs mass term $\mu$.

What has been learned is that $t-b-\tau$ Yukawa coupling unification {\it does} occur in the 
MSSM for $\mu >0$ (as preferred by the $(g-2)_\mu$ anomaly), 
but {\it only if certain conditions} are satisfied.
\bi
\item The scalar mass parameter $m_{16}$ should be very heavy: in the range 5-20 TeV.
\item The gaugino mass parameter $m_{1/2}$ should be as small as possible.
\item The SSB terms should be related as $A_0^2=2m_{10}^2=4m_{16}^2$, with
$A_0=-2m_{16}$ (in our sign convention). This combination
was found to yield a radiatively induced inverted scalar mass hierarchy (IMH) by Bagger
{\it et al.}\cite{bfpz} for MSSM+right hand neutrino (RHN) models with Yukawa coupling unification.
\item $\tan\beta \sim 50$.
\item EWSB can be reconciled with Yukawa unification only if the Higgs SSB masses
are split at $M_{GUT}$ such that $m_{H_u}^2 <m_{H_d}^2$. The HS prescription ends up
working better than DT splitting\cite{bdr,bf}. 
\ei

In the case where the above conditions are satisfied, then Yukawa coupling unification to within 
a few percent can be achieved. The resulting sparticle mass spectrum has some notable 
features.
\bi
\item First and second generation matter scalars have masses of order $m_{16}\sim 5-20$ TeV.
\item Third generation scalars, $m_A$ and $\mu$ are suppressed relative to $m_{16}$
by the IMH mechanism: they have masses on the $1-2$ TeV scale. This reduces the amount of 
fine-tuning one might otherwise expect in such models.
\item Gaugino masses are quite light, with $m_{\tg}\sim 350-500$ GeV, 
$m_{\tz_1}\sim 50-80$ GeV and $m_{\tw_1}\sim 100-150$ GeV.
\ei

The sparticle mass spectra from $SO(10)$ SUSY GUTs shares some features with spectra
generated in ``large cutoff supergravity'' or LCSUGRA, investigated in Ref. \cite{lcsugra}.
LCSUGRA also has high mass scalars-- typically with mass around 5 TeV-- and low mass 
gauginos. The $SO(10)$ SUSY GUT models are different from LCSUGRA in that they 
have a large $A_0$, with $A_0\sim -2m_{16}$, and a $\mu$ term of around 1-2 TeV.
This means $SO(10)$ SUSY GUTs have a dominantly bino-like $\tz_1$ state, whereas
the LCSUGRA authors adopt the mSUGRA model focus point region, 
which has a mixed higgsino-bino $\tz_1$ state. The latter can easily give the
measured abundance of cold dark matter (CDM) in the form of lightest neutralinos.

Since the lightest neutralino of $SO(10)$ SUSY GUTs is nearly a pure bino state, it turns out the 
neutralino relic density $\Omega_{\tz_1}h^2$ is calculated to be extremely high, 
of order $10^2-10^4$. This conflicts with the WMAP-measured value\cite{wmap}, which gives
\be
\Omega_{CDM}h^2\equiv \rho_{CDM}/\rho_c =
0.111^{+0.011}_{-0.015}\ \ (2\sigma ) .
\label{eq:Oh2}
\ee
where $h=0.74\pm 0.03$ is the scaled Hubble constant.

Several solutions to the $SO(10)$ SUSY GUT dark matter problem have been proposed in
Refs. \cite{abbk,bkss}. Here, we will concentrate on the most attractive one: that
the dark matter particle is in fact not the neutralino, but the {\it axino} $\ta$. 
Axino dark matter occurs in models where the MSSM is extended via the Peccei-Quinn (PQ)
solution to the strong $CP$ problem\cite{nr}. The PQ solution introduces a spin-0 axion field
into the model; if the model is supersymmetric, then a spin-${1\over 2}$ axino
is also required. It has been shown that the $\ta$ state can be an excellent 
candidate for cold dark matter in the universe\cite{axino}. In this paper, we will find that
$SO(10)$ SUSY GUT models with an axino DM candidate can 1. yield the correct
abundance of CDM in the universe, 2. avoid the gravitino/BBN problem and 3. have an compelling
mechanism for generating the matter-antimatter asymmmetry of the universe via 
non-thermal leptogenesis.

\section{The gravitino problem}
\label{sec:gravitino}

An affliction common to all models with gravity mediated SUSY breaking 
(supergravity or SUGRA) models is known as the gravitino problem. In 
realistic SUGRA models (those that include the SM as their sub-weak-scale effective
theory), SUGRA is broken in a hidden sector by the superHiggs mechanism, 
which induces a mass for the gravitino $\tG$, which is commonly taken to be of 
order the weak scale. The gravitino mass $m_{\tG}$ ends up setting the mass scale
for all the soft breaking terms, so then all SSB terms end up also being of order
the weak scale.

The coupling of the gravitino to matter is strongly suppressed by the Planck 
mass, so the $\tG$ in the mass range considered here ($m_{\tG}\sim 5-20$ TeV) 
is never in thermal equilibrium with the thermal bath in 
the early universe. Nonetheless, it does get produced by scatterings of
particles that do partake of thermal equilibrium.
Thermal production of gravitinos in the early universe has been calculated 
in Refs. \cite{relic_G}, where the abundance is found to depend naturally on 
$m_{\tG}$ and on the re-heat temperature $T_R$ at the end of inflation.
Once produced, the $\tG$s decay into all varieties of particle-sparticle
pairs, but with a lifetime that can exceed $\sim 1$ sec, the time scale
where Big Bang nucleosynthesis (BBN) begins. 
The energy injection from $\tG$ decays is a threat to dis-associate the
light element nuclei which are created in BBN.
Thus, the long-lived $\tG$s can destroy the successful predictions of the 
light element abundances as calculated by nuclear thermodynamics.

The BBN constraints on gravitino production in the early universe have been
calculated by several groups\cite{bbn_gino}. The recent 
results from Ref. \cite{Kohri} give an upper limit on the re-heat temperature
as a function of $m_{\tG}$. The results depend on how long-lived the $\tG$
is (at what stage of BBN the energy is injected), and what its dominant decay
modes are. Qualitatively, for $m_{\tG}\alt 5$ TeV, $T_R\alt 10^6$ GeV
is required; if this is violated, then too many $\tG$ are produced in the early
universe, which detroy the $^3He$, $^6Li$ and $D$ abundance calculations.
For $m_{\tG}\sim 5-50$ TeV, the re-heat upper bound is much less:
$T_R\alt 5\times 10^7 -10^9$ GeV (depending on the $^4He$ abundance) 
due to overproduction of $^4He$ arising from
$n\leftrightarrow p$ conversions. For $m_{\tG}\agt 50$ TeV, there is an upper bound 
of $T_R\alt 5\times 10^9$ GeV due to overproduction of $\tz_1$ LSPs due to 
$\tG$ decays. 

Solutions to the gravitino BBN problem then include: 1. having $m_{\tG}\agt 50$ TeV
but with an unstable $\tz_1$ (no $T_R$ bound),
2. having a gravitino LSP so that $\tG$ is stable or 3. keep the re-heat
temperature below the BBN bounds. We will here adopt solution number 3. 
In the case of $SO(10)$ SUSY GUT models, with $m_{\tG}\sim m_{16}\sim 5-20$ TeV, 
this means we need a re-heat temperature $T_R\alt 10^8 - 10^9$ GeV. 

\section{Non-thermal leptogenesis}
\label{sec:lepto}

The data gleaned on neutrino masses during the past decade has lead credence to 
a particular mechanism of generating the baryon asymmetry of the universe 
known as leptogenesis\cite{THlepto}. Leptogenesis requires the presence of heavy gauge singlet
Majorana right handed neutrino states $\psi_{N^c_i}(\equiv N_i)$ with mass $M_{N_i}$
($i=1-3$ is a generation index). 
The $N_i$ states may be produced thermally in the early universe, or perhaps 
non-thermally, as suggested in Ref. \cite{NTlepto} via inflaton $\phi \to N_iN_i$
decay. The $N_i$ may then decay asymmetrically to elements of the doublets-- 
for instance $\Gamma (N_1\to h_u^+ e^-)\ne \Gamma (N_1\to h_u^- e^+)$-- owing
to the contribution of $CP$ violating phases in the tree/loop decay
interference terms. Focussing on just one species of heavy neutrino $N_1$, 
the asymmetry is calculated to be\cite{epsilon}
\be
\epsilon\equiv \frac{\Gamma (N_1\to \ell^+)-\Gamma (N_1\to \ell^-)}{\Gamma_{N_1}}
\simeq -\frac{3}{8\pi}\frac{M_{N_1}}{v_u^2}m_{\nu_3}\delta_{eff} ,
\ee 
where $m_{\nu_3}$ is the heaviest active neutrino, $v_u$ is the up-Higgs vev and
$\delta_{eff}$ is an effective $CP$-violating phase factor which may be of order 1.
The ultimate baryon asymmetry of the universe is proportional to $\epsilon$, 
so larger values of $M_{N_1}$ lead to a higher baryon asymmetry. 

To find the baryon asymmetry, one may first assume that the $N_1$ is thermally 
produced in the early universe, and then solve the Boltzmann equations for the $B-L$
asymmetry. The ultimate baryon asymmetry of the universe arises from the lepton asymmetry
via sphaleron effects. The final answer\cite{bp}, compared against 
the WMAP-measured result $\frac{n_B}{s}\simeq 0.9\times 10^{-10}$ for the baryon-to-entropy
ratio, requires $M_{N_1}\agt 10^{10}$ GeV, and thus a re-heat temperature $T_R\agt 10^{10}$ GeV.
This high a value of reheat temperature is in conflict with the upper bound on $T_R$
discussed in Sec. \ref{sec:gravitino}. 
In this way, it is found that generic SUGRA models are apparently in conflict with
leptogenesis as a means to generate the baryon asymmetry of the universe.

If one instead looks to non-thermal leptogenesis, then it is possible to have lower
reheat temperatures, since the $N_1$ may be generated via inflaton decay. 
The Boltzmann equations for the $B-L$ asymmetry have been solved numerically in Ref. \cite{imy}.
The $B-L$ asymmetry is then converted to a baryon asymmetry via sphaleron effects
as usual. 
The baryon-to-entropy ratio is calculated in \cite{imy}, where it is found
\be
\frac{n_B}{s}\simeq 8.2\times 10^{-11}\times \left(\frac{T_R}{10^6\ {\rm GeV}}\right) 
\left(\frac{2M_{N_1}}{m_\phi}\right) \left(\frac{m_{\nu_3}}{0.05\ {\rm eV}}\right) \delta_{eff} ,
\ee
where $m_\phi$ is the inflaton mass.
Comparing calculation with data, a lower bound $T_R\agt 10^6$ GeV may be inferred
for viable non-thermal leptogenesis via inflaton decay. 

\section{Axino dark matter}
\label{sec:axino}

The sparticle mass spectrum described in Sec. \ref{sec:so10} is characterized
by $5-20$ GeV scalars, but very light gauginos, with a $\mu$ parameter
of order 1-2 TeV. As a consequence, the neutralino $\tz_1$ ends up being 
nearly pure bino. Since all the scalars are quite heavy, the predicted
neutralino relic abundance ends up being very high: the calculation 
of Refs. \cite{abbk,bkss} find values in the range
$\Omega_{\tz_1}h^2\sim 10^2 -10^4$, which is $3-4$ orders of magnitude
beyond the WMAP-measured abundance. 

A solution was advocated in Ref. \cite{bkss} that in fact the $\tz_1$
state is not the LSP, but instead the {\it axino} $\ta$ makes up the
CDM of the universe. The axino is the spin-${1/2}$ element of the
axion supermultiplet which is needed to solve the strong $CP$ 
problem in supersymmetric models. The axino is characterized by a 
mass in the range of keV-GeV. Its couplings are of sub-weak interaction 
strength, since they are suppressed by the Peccei-Quinn
symmetry breaking scale $f_a$, which itself has a viable mass range
$10^{10}-10^{12}$ GeV. While the axino interacts very feebly, it does interact
more strongly than the gravitino. 

If the $\ta$ is the lightest SUSY particle, then the $\tz_1$ will no longer
be stable, and can decay via $\tz_1\to \ta\gamma$. 
The relic abundance of axinos from neutralino decay 
(non-thermal production, or $NTP$) is given simply by
\be
\Omega_{\ta}^{NTP}h^2 =\frac{m_{\ta}}{m_{\tz_1}}\Omega_{\tz_1}h^2 ,
\label{eq:Oh2_NTP}
\ee
since in this case the axinos inherit the thermally produced 
neutralino number density.
Notice that neutralino-to-axino decay offers a mechanism to shed
large factors of relic density. For a case where $m_{\tz_1}\sim 50$
GeV and $\Omega_{\tz_1}h^2\sim 1000$, as can occur in $SO(10)$ SUSY GUTs,
an axino mass of less than 5 MeV reduces the DM abundance to below
WMAP-measured levels.

The lifetime for these decays has been calculated, 
and it is typically in the range of $\tau (\tz_1\to \ta\gamma )\sim 0.03$ sec\cite{axino}. 
The photon energy injection from $\tz_1\to\ta\gamma$ decay 
into the cosmic soup occurs well before
BBN, thus avoiding the constraints that plague the case of a gravitino LSP\cite{feng}.
The axino DM arising from neutralino decay is generally 
considered warm or even hot dark matter for cases with 
$m_{\ta}\alt 1-10$ GeV\cite{jlm}.
Thus, in our Yukawa-unified scenario, where $m_{\ta}\alt 80$ MeV, we {\it always}
get warm DM from neutralino decay.

Even though they are not in thermal equilibrium, axinos can still
be produced thermally in the early universe via scattering processes.
The axino thermally produced (TP) relic abundance has been 
calculated in Ref. \cite{axino,steffen}, and is given by
\be
\Omega_{\ta}^{TP}h^2\simeq 5.5 g_s^6\ln\left(\frac{1.108}{g_s}\right)
\left(\frac{10^{11}\ {\rm GeV}}{f_a/N}\right)^2 
\left(\frac{m_{\ta}}{0.1\ {\rm GeV}}\right) 
\left(\frac{T_R}{10^4\ {\rm GeV}}\right)
\label{eq:Oh2_TP}
\ee
where $g_s$ is the strong coupling evaluated at $Q=T_R$ and $N$ is the
model dependent color anomaly of the PQ symmetry, of order 1.
The thermally produced axinos qualify as {\it cold} dark matter as long as 
$m_{\ta}\agt 0.1$ MeV\cite{axino,steffen}.

\section{A consistent cosmology for axino DM from $SO(10)$ SUSY GUTs}
\label{sec:results}

At this point, we are able to check if we can implement a consistent cosmology
for $SO(10)$ SUSY GUTs with axino dark matter.
Our first step is to select points from the $SO(10)$ parameter space Eq. \ref{eq:pspace} 
that are very nearly Yukawa-unified.
In Ref. \cite{bkss}, Yukawa unified solutions
were searched for by looking for $R$ values as close to 1 as possible, where 
\be
R=\frac{max(f_t,\ f_b,\ f_\tau )}{min(f_t,\ f_b,\ f_\tau )}
\ee
where the $f_t$, $f_b$ and $f_\tau$ Yukawa couplings were evaluated
at $M_{GUT}$. Thus, a solution with $R=1.05$ gives Yukawa unification to 5\%.

We would like solutions where the axino DM is dominantly CDM. 
For definiteness, we will insist on $\Omega_{\ta}^{NTP}h^2\sim 0.01$, 
while $\Omega_{\ta}^{TP}h^2=0.1$. 
Thus, in step 1., we select models from the random scan of Ref. \cite{bkss} that
have $R<1.05$, and $m_{16}:5-20$ TeV. 
In step 2., from the known value of $m_{\tz_1}$ and $\Omega_{\tz_1}h^2$, 
we next calculate the axino mass needed to generate $\Omega_{\ta}^{NTP}h^2=0.01$
according to Eq. \ref{eq:Oh2_NTP}.
In step 3, we plug $m_{\ta}$ into Eq. \ref{eq:Oh2_TP}, where we also take
$g_s=0.915$ (the running $g_s$ value at $\sim 10^6$ GeV), and PQ scale
$f_a/N= 10^{12}$ GeV. By insisting that $\Omega_{\ta}^{TP}h^2=0.1$, we may calculate the
value of $T_R$ that is needed.

\begin{figure}[!t]
\begin{center}
\epsfig{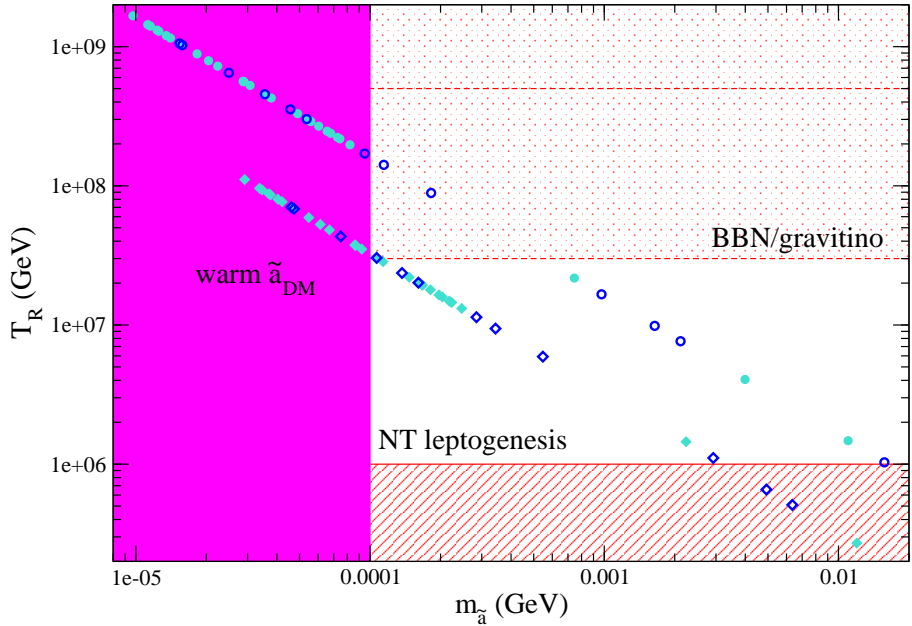}
\end{center}
\caption{\small\it 
Plot of Yukawa unified solutions with $R<1.05$ and 5 TeV $<m_{16}<$ 20 TeV 
in the $m_{\ta}\ vs. T_R$ plane. The upper band of solutions has
$\Omega_{\ta}^{NTP}h^2=0.01$, $\Omega_{\ta}^{TP}h^2=0.10$ and $f_a/N=10^{12}$ GeV, 
while the lower band of solutions has $\Omega_{\ta}^{NTP}h^2=0.03$, $\Omega_{\ta}^{TP}h^2=0.08$ 
and $f_a/N=5\times 10^{11}$ GeV.
}
\label{fig:mavsTR} 
\end{figure}

Our results are plotted in the $m_{\ta}\ vs.\ T_R$ plane in 
Fig. \ref{fig:mavsTR}, and occupy the upper band of solutions. 
In this plane, solutions with $T_R\alt 3\times 10^7-5\times 10^8$ GeV 
are allowed by the gravitino constraint (with $m_{\tG}\sim 5-20$ TeV) and BBN. 
Solutions with $T_R\agt 10^6$ GeV
can generate the matter-antimatter asymmetry correctly via non-thermal
leptogenesis. Solutions with $m_{\ta}\agt 10^{-4}$ GeV give dominantly {\it cold} DM
from TP of axinos. 
Solutions with $m_{16}>15$ TeV are denoted by filled (turquoise) symbols, while solutions with
$m_{16}<15$ TeV have open (dark blue) symbols.

We see that a variety of points fall in the allowed region. These points
give rise to a {\it consistent cosmology} for $SO(10)$ SUSY GUT models!
Of course, there is some uncertainty in these results. We can take higher or lower
values of the PQ breaking scale, higher or lower fractions of 
$\Omega_{\ta}^{NTP}h^2$, and
the $T_R$ upper (and lower) bounds have some variability built into them. 
As an example, the lower band of solutions is obtained with 
$\Omega_{\ta}^{NTP}h^2=0.03$,
$\Omega_{\ta}^{TP}h^2=0.08$ and $f_a/N =5\times 10^{11}$ GeV. In this case, some of the
previously excluded solutions migrate into the allowed region to give a consistent
cosmology with somewhat different parameters.

\section{Conclusion}

Our main conclusion can be summarized briefly.
For Yukawa unified supersymmetric models, as expected in $SO(10)$ SUSY GUT models, 
we find one can implement a consistent cosmology including the following:
1. BBN safe mass spectra owing to the multi-TeV value of $m_{16}$, which arises
in SUGRA models from a multi-TeV $m_{\tG}$
2. a WMAP-allowed relic density of CDM that consists dominantly of
thermally produced axinos, and 3. the re-heat temperature needed to fulfill
the relic density falls {\it above} the lower bound required by non-thermal leptogenesis, 
and {\it below} the upper bound coming from gravitino/BBN constraints.

We feel that the fact that Yukawa unified $SO(10)$ SUSY GUT
models pass these several cosmological tests makes them even more compelling
than they were based on pure particle physics reasons. In any case, with
a spectrum of light gluinos, charginos and neutralinos, they should easily be tested
by experiments at the CERN LHC\cite{etmiss} even with low integrated luminosities of
just $\sim 0.1$ fb$^{-1}$.

\section*{Acknowledgments} HB would like to thank 
W. Buchm\"uller and V. Barger for conversations.
This research was supported in part by the National Science Foundation under
Grant No. NSF PHY05-51164.
%

\end{document}